\documentclass{article}
\usepackage{emulateapj,pstricks,apjfonts}
\def\chandra	{{\em Chandra}\/}
\def\xmm	{{\em XMM}\/}
\def\gax	{\gtrsim}
\def\lax	{\lesssim}
\def\cmsq	{~cm$^{-2}$}
\def\cmcube	{~cm$^{-3}$}
\def\deg        {$^{\circ}$}
\begin{document}

\submitted{ApJ Letters in press}

\lefthead{X-RAY GUNN-PETERSON TEST}
\righthead{MARKEVITCH}

\title{A DIFFERENTIAL X-RAY GUNN-PETERSON TEST USING A GIANT CLUSTER
FILAMENT}

\author{Maxim Markevitch}
\affil{Harvard-Smithsonian Center for Astrohysics, 60 Garden St.,
Cambridge, MA 02138; maxim@head-cfa.harvard.edu}

\begin{abstract}

Using CCD detectors onboard the forthcoming X-ray observatories \chandra\
and \xmm, it is possible to devise a measurement of the absolute density of
heavy elements in the hypothetical warm gas filling intercluster space. This
gas may be the largest reservoir of baryonic matter in the Universe, but
even its existence has not been proven observationally at low redshifts. The
proposed measurement would make use of a unique filament of galaxy clusters
spanning over $700\,h^{-1}_{50}$ Mpc ($0.1\lax z \lax 0.2$) along the line
of sight in a small area of the sky in Aquarius. The surface density of
Abell clusters there is more than 6 times the sky average. It is likely that
the intercluster matter column density is enhanced by a similar factor,
making its detection feasible under certain optimistic assumptions about its
density and elemental abundances. One can compare photoabsorption depth,
mostly in the partially ionized oxygen edges, in the spectra of clusters at
different distances along the filament, looking for a systematic increase of
depth with the distance. The absorption can be measured by the same detector
and through the same Galactic column, hence the differential test.  A CCD
moderate energy resolution ($\Delta E\sim 100$ eV) is adequate for detecting
an absorption edge at a known redshift.

\end{abstract}


\section{INTRODUCTION}

Simulations suggest that at present, most of the baryons in the Universe
should reside in the diffuse medium filling intergalactic space (e.g.,
Miralda-Escud\'e et al.\ 1996; Cen \& Ostriker 1999a). Indeed, the observed
baryonic matter in low-redshift galaxies and clusters is only a fraction of
the amount predicted by Big Bang nucleosynthesis (Persic \& Salucci 1992;
Fukugita, Hogan, \& Peebles 1998). Observationally, however, little is known
about intergalactic medium (IGM) outside the relatively small confines of
galaxy clusters where it is sufficiently hot to emit detectable radiation.
UV and optical studies show that intercluster hydrogen is almost completely
ionized and therefore practically undetectable (Gunn \& Peterson 1965), with
only a small fraction of it residing in Ly$\alpha$ forest clouds (e.g.,
Rauch et al.\ 1997; Giallongo, Fontana, \& Madau 1997). IGM may be enriched
with heavy elements that originate in supernovae and are transported to IGM
by supernova-generated winds (e.g., De Young 1978). The amount of heavy
elements in the IGM carries important information on the cumulative number
of supernova explosions integrated over time, which is inaccessible by other
means. Heavy elements with a relative abundance of order 1\% solar were
detected in Ly$\alpha$ absorbers (e.g., Burles \& Tytler 1996 and references
therein). However, these measurements are restricted by design to the small
fraction of high-$z$ gas that is in dense clouds, and are unlikely to tell
about the IGM on average.

At present, there is no direct information on the amount and composition of
the diffuse IGM at low $z$ --- possibly the largest reservoir of baryons
around us --- although there are plausible conjectures. At low redshifts,
gas in galaxies has roughly solar heavy element abundances and gas in
clusters exhibits quite a universal relative iron abundance of 1/3 solar
(e.g., Edge \& Stewart 1991; Fukazawa et al.\ 1998). The apparent lack of
iron abundance evolution in clusters out to $z\sim 0.3$ and possibly beyond
puts the epoch of cluster gas enrichment at $z\gax 1$ (Mushotzky \&
Loewenstein 1997).  Today's clusters should have formed later than that, and
there is no obvious reason for the enrichment of IGM to occur preferentially
in the regions of space that later became clusters.  Therefore, as argued
by, e.g., Renzini (1999), cluster metallicity can be taken as representative
of the low-$z$ universe as a whole.  Simulations by Cen \& Ostriker (1999b)
predict a present-day universal abundance of 0.2 solar with a higher metal
concentration around clusters. On the other hand, clusters, unlike the field
population, contain mostly ellipticals and few spirals, suggesting that the
dense cluster environment provides for an effective stripping of the
enriched galaxy gas (e.g., Sarazin 1988). If so, the IGM may have much lower
heavy element abundances than the cluster gas.  Note that if heavy elements
are detected in intercluster space, this would also set a lower limit on the
total hydrogen density (and therefore the baryon density $\Omega_b$) as
well, since the intercluster relative abundances are unlikely to be higher
than the cluster values.

Unlike hydrogen and most helium, heavy elements in the diffuse IGM outside
clusters are not expected to be strongly ionized and can in principle be
detected via X-ray absorption in an X-ray analog of the Gunn-Peterson test
(Shapiro \& Bahcall 1980; Aldcroft et al.\ 1994; Fang \& Canizares 1997;
Perna \& Loeb 1998; Hellsten, Gnedin, \& Miralda-Escud\'e 1998). Such an
absorption (mostly due to oxygen and iron) has never been observed. Indeed,
the above authors estimate that for an IGM that is uniform on large linear
scales, the expected resonant absorption lines in a random direction towards
a distant quasar are very weak.

Fortunately, the Universe is not uniform and there is one place in the sky
where such a test, using not quasars but galaxy clusters as background
candles, may be feasible with the forthcoming \chandra\ and \xmm\
observatories, as described below.

%%%%%%%%%%%%%%%%%%%%%%%%%%%%%%%%%%%%%%%%%%%%%%%%%%%%%%%%%%%%%%%%%%%%%%%%%%
\begin{figure*}[b]
\pspicture(0,0.5)(18.5,9)
%\psgrid(0,1.4)(16.5,9)

\rput[tl]{0}(0.4,9.7){\epsfxsize=8.5cm
\epsffile{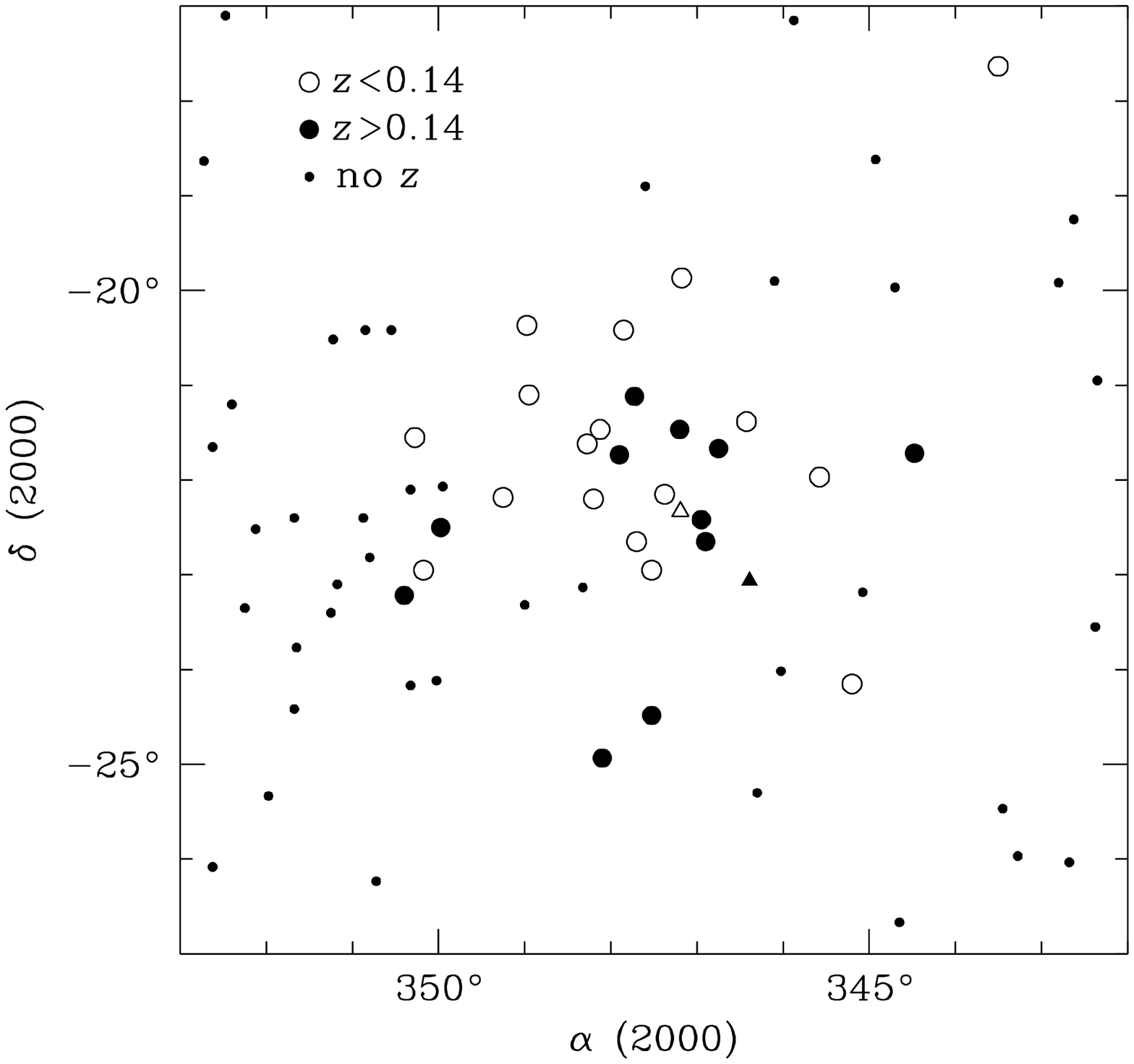}}

\rput[tl]{0}(9.5,9.7){\epsfxsize=8.5cm
\epsffile{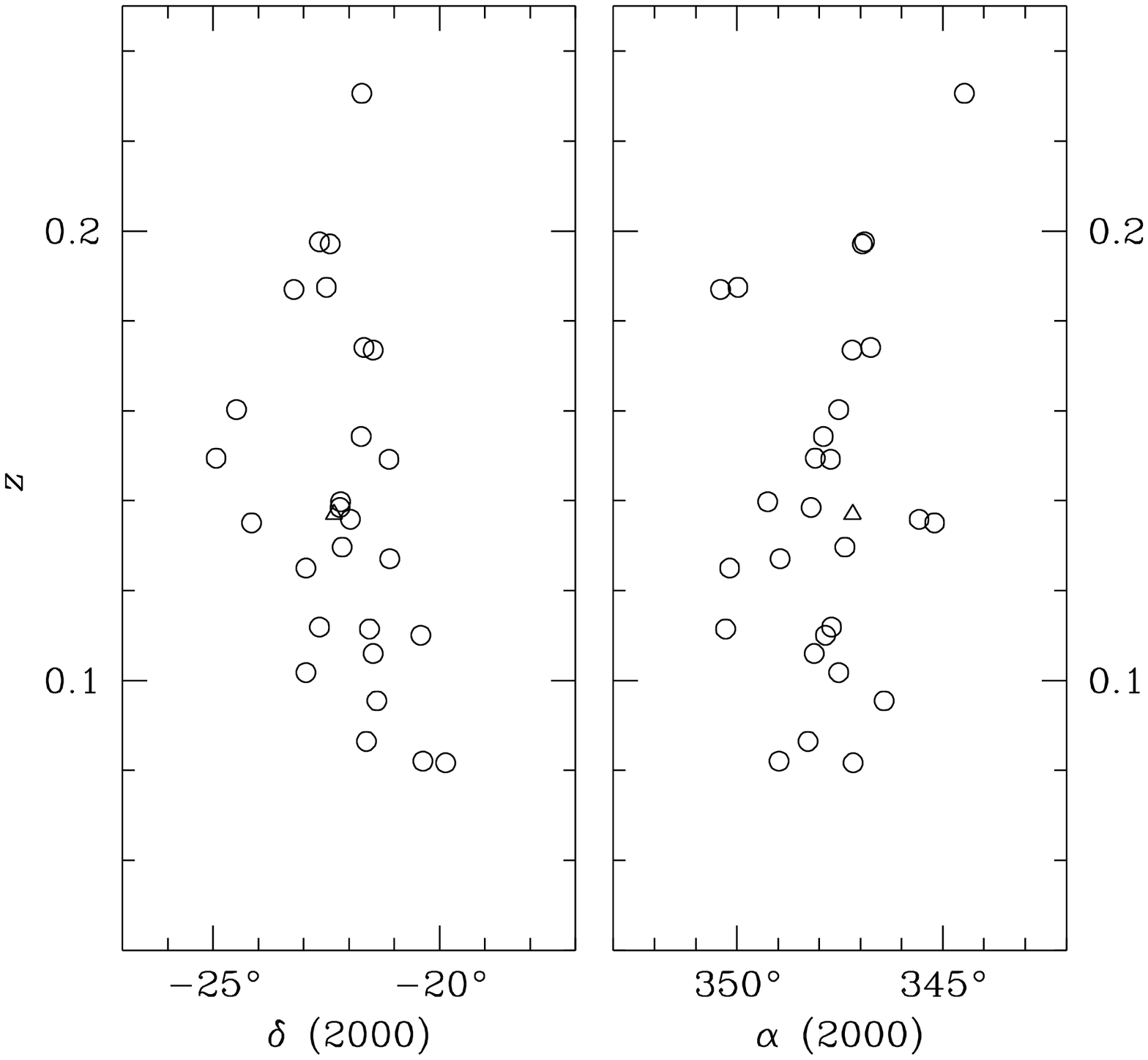}}

\rput[bl]{0}(7.5,2.8) {\large\it a}
\rput[bl]{0}(13.5,2.8){\large\it b}
\rput[bl]{0}(16.9,2.8){\large\it c}

\rput[tl]{0}(0,1.4){
\begin{minipage}{18.5cm}
\small\parindent=3.5mm
{\sc Fig.}~2.---({\em a}) Area of the cluster filament enlarged. 
Clusters beyond the median redshift are shown as big black symbols, the
nearer ones shown as open symbols and the ones with unknown $z$ as small
symbols. Triangles mark two QSOs. ({\em b,c}) Two redshift projections of
the area in panel ({\em a}), showing only clusters with known $z$. The scale
of the $z$ axis is arbitrary.
\par
\end{minipage}
}
\endpspicture
\end{figure*}
%%%%%%%%%%%%%%%%%%%%%%%%%%%%%%%%%%%%%%%%%%%%%%%%%%%%%%%%%%%%%%%%%%%%%%%%%%

\section{LOOKING THROUGH A GIANT CLUSTER FILAMENT}

Figure~1 shows the sky distribution of the most distant galaxy clusters from
the catalog by Abell, Corwin, \& Olowin (1989). Besides the Galactic plane
shadow, the most prominent feature of the distribution is a concentration of
clusters in Aquarius at $\alpha=348$\deg, $\delta=-23$\deg, with the surface
density of clusters about 6 times higher than average over the Southern
Galactic hemisphere. This concentration was first noted by Abell (1961) and
listed among the most likely candidates of rich distant superclusters. Later
spectroscopic data (Ciardullo, Ford, \& Harms 1985) showed that members of
this concentration in fact form a giant filament along the line of sight,
spanning a range of redshifts between 0.08 and 0.21 or about
700\,$h^{-1}_{50}$~Mpc (Fig.\ 2). One can reasonably assume that the
overdensity of the IGM in this volume is proportional to that of clusters,
since a bias on such a large linear scale is unlikely (forthcoming
cosmological simulations of the Hubble volume will address this issue, e.g.,
Colberg et al.\ 1998). Therefore, this filament should also provide an
enhancement in the IGM column density by a factor of about 6. This
enhancement makes the absorption detectable in the X-ray spectra of clusters
located at the far end of the filament that are seen through the filament.

%%%%%%%%%%%%%%%%%%%%%%%%%%%%%%%%%%%%%%%%%%%%%%%%%%%%%%%%%%%%%%%%%%%%%%%%%%
\noindent
%\begin{figure*}[tb]
\pspicture(0,0.2)(9,7.5)
%\psgrid(0,0)(15,7.2)

\rput[tl]{0}(0,7){\epsfxsize=8.8cm
\epsffile[69 300 570 565]{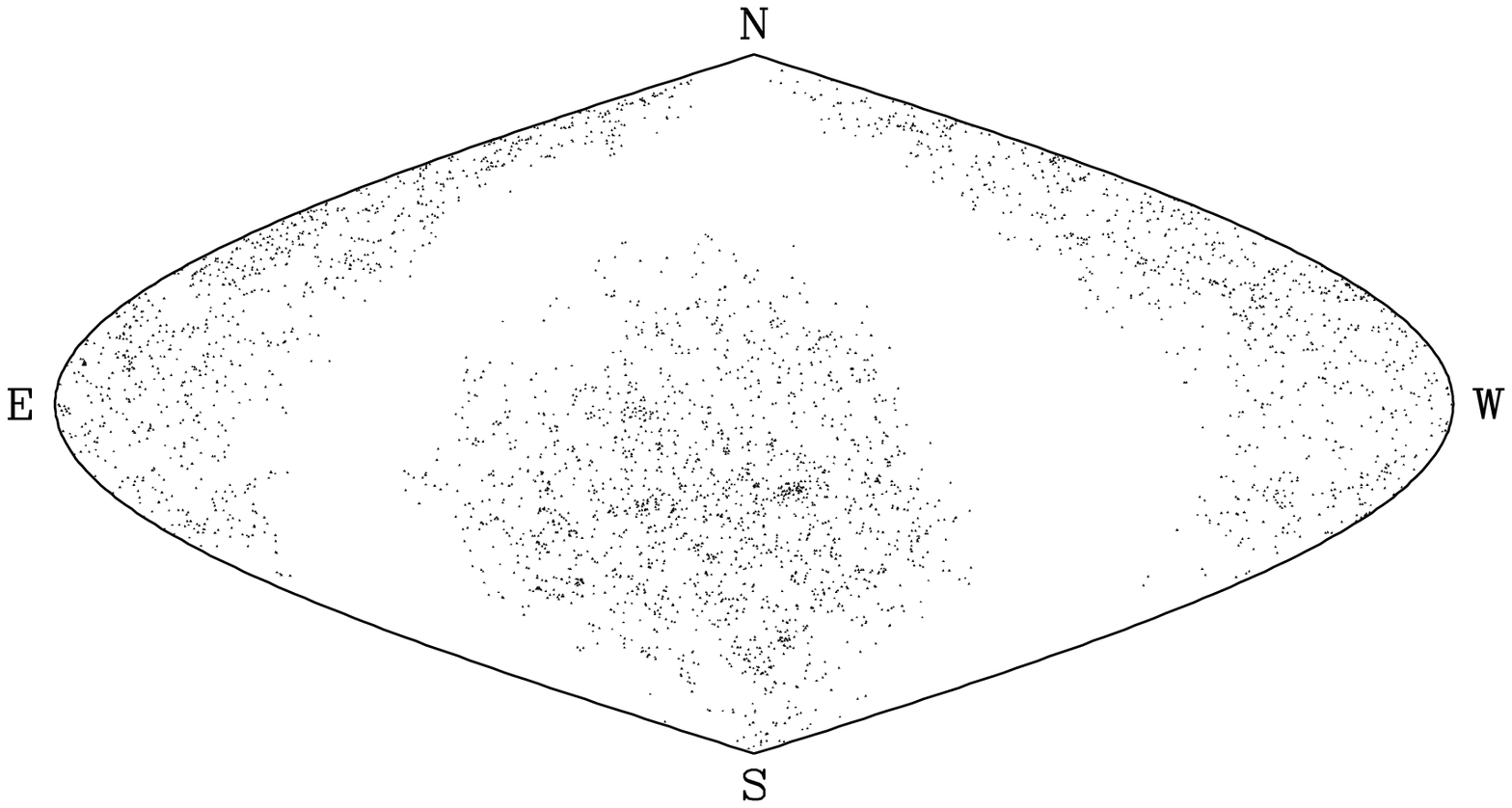}}

\rput[tl]{0}(0,1.8){
\begin{minipage}{8.75cm}
\small\parindent=3.5mm
{\sc Fig.}~1.---Abell distance class 5 clusters plotted in equatorial
coordinates (the center corresponds to $\alpha=0,\, \delta=0$). The most
prominent cluster concentration near the center is shown enlarged in Fig.\
2.
\par
\end{minipage}
}
\endpspicture
%\end{figure*}
%%%%%%%%%%%%%%%%%%%%%%%%%%%%%%%%%%%%%%%%%%%%%%%%%%%%%%%%%%%%%%%%%%%%%%%%%%

There are distant and nearby members of the filament that are close in
projection (Fig.\ 2). This enables a ``differential'' test by measuring a
difference in absorption in the nearby and distant cluster spectra and
looking for its systematic increase with the cluster distance along the
filament.  Such a differential measurement can significantly reduce
systematic errors due to (a) uncertainty of the Galactic column density and
(b) unavoidable instrument calibration inaccuracy. Since the hypothetical
absorbing gas is located within the small interval of known and low
redshifts, its detection and interpretation can be straightforward.

\subsection{Expected absorption column density}

Big Bang nucleosynthesis predicts a baryonic (plausibly, dominated by IGM)
density parameter of $\Omega_b \simeq 0.05\, h_{50}^{-2}$ (Walker et al.\
1991) $\pm$ a factor of 2 from the uncertainly of the measured deuterium
abundance (e.g., Steigman 1996). Simulations of the observed Ly$\alpha$
forest by Rauch et al.\ (1997) and the detection of high-$z$ helium by
Davidsen, Krauss, \& Wei (1996) are consistent with the upper bound of the
above value. Here we would like to get an upper limit estimate of the
possible absorbing column, for which we assume the above upper bound as an
IGM density, $\Omega_{\rm IGM}\simeq 0.1\,h_{50}^{-2}$. The difference in
hydrogen column density between the nearest and farthest clusters in the
filament is
\begin{equation}
N_H\,(0.1<z<0.2)\approx 2.5\times 10^{21}\; {\rm cm}^{-2} \,h^{-1}_{50}\;
\frac{\Omega_{\rm IGM}}{0.1\,h^{-2}_{50}}\;\frac{\Delta}{6}\;\frac{1}{b_{700}},
\end{equation}
where $\Delta$ is the surface overdensity of clusters in that region of the
sky and $b_{700}$ is the cluster bias on the 700 Mpc scale (if there is
any). For comparison, the Galaxy has $N_H=2\times 10^{20}$\cmsq\ in that
direction (of course, most of the Galactic absorption is due to neutral
hydrogen absent in IGM). In the redshift interval in front of the filament
($z<0.1$) and assuming no IGM overdensity, $N_H\approx 2\times
10^{20}$\cmsq, negligible compared to that in the filament. Another useful
quantity for comparison is a possible column density of warm gas on a line
of sight crossing the outskirts of a rich galaxy cluster. For example,
extrapolating the X-ray gas density profile for the Coma cluster beyond its
virial radius ($r\approx 3$ Mpc) to 10 Mpc, one obtains $N_H\sim 3\times
10^{20}$\cmsq\ (excluding the hot, weakly absorbing gas within the virial
radius). Again, it is negligible compared to the column density that
accumulates along the 700 Mpc overdense filament. Thus the proposed test can
indeed probe the absorbing medium far from cluster confines.

\subsection{Expected conditions in the gas} 

Simulations predict that a large fraction of all baryons at low redshifts is
in the form of ``warm'' gas with temperatures between $10^5$ and $10^7$~K
(e.g., Cen \& Ostriker 1999a). Although these temperatures are low for a
complete collisional ionization of heavy elements such as oxygen and iron,
in the low-density regions outside clusters, photoionization by Cosmic X-ray
Background (CXB) dominates the ionization rate (e.g., Aldcroft et al.\
1994). Unlike the uncertain ionizing UV background at high redshifts that is
involved in the interpretation of Ly$\alpha$ data, the present-day CXB is
directly measurable (e.g., Chen, Fabian, \& Gendreau 1997). The ionizing
radiation is stronger in the immediate vicinity of an X-ray-bright cluster,
but CXB dominates beyond a few Megaparsecs from the cluster. The diffuse IGM
is optically thin with respect to photoabsorption. For the expected density,
the recombination timescales of interest are short compared to the Hubble
time, thus the medium should be close to ionization equilibrium. For an
estimate of the ionization balance expected in the gas filling our filament
and subjected to photoionization by CXB, we used XSTAR code by T. Kallman
and J.  Krolik%
\footnote{ftp://legacy.gsfc.nasa.gov/software/plasma\_codes/xstar}.
For a range of temperatures $(3-30)\times 10^5$~K and an IGM density
$n_H\approx 2\times 10^{-6}$\cmcube\ (six times the assumed $\Omega_{\rm
IGM}$), oxygen is mainly in the form of O{\small VII} and O{\small VIII}
ions with an increasing fraction of the completely ionized species for
increasing temperature. If the IGM is clumpy (which is most likely, e.g.,
Cen \& Ostriker 1999a), plasma in the denser regions would have lower
ionization states, for a given temperature. For illustrative purposes of
this paper, we use the ionization balance calculated for $T=3\times 10^5$~K
and the above average gas density; qualitative conclusions are similar for
other temperatures and densities in the expected range. For a consistently
optimistic estimate, relative abundances of heavy elements in the IGM are
assumed to be 0.3 solar. To calculate absorption depth for the obtained
mixture of ions, atomic data from Verner et al.\ (1996ab) were used. Figure
3{\em a} shows an example absorbed spectrum.  As noted in earlier works, of
all elements, oxygen causes the strongest absorption features and has the
best chance to be detected.

%%%%%%%%%%%%%%%%%%%%%%%%%%%%%%%%%%%%%%%%%%%%%%%%%%%%%%%%%%%%%%%%%%%
\begin{figure*}[tb]
\pspicture(0,10.4)(18.5,19.7)
%\psgrid(0,0)(18.5,20)

\rput[tl]{0}(0.4,20.7){\epsfxsize=8.5cm
\epsffile{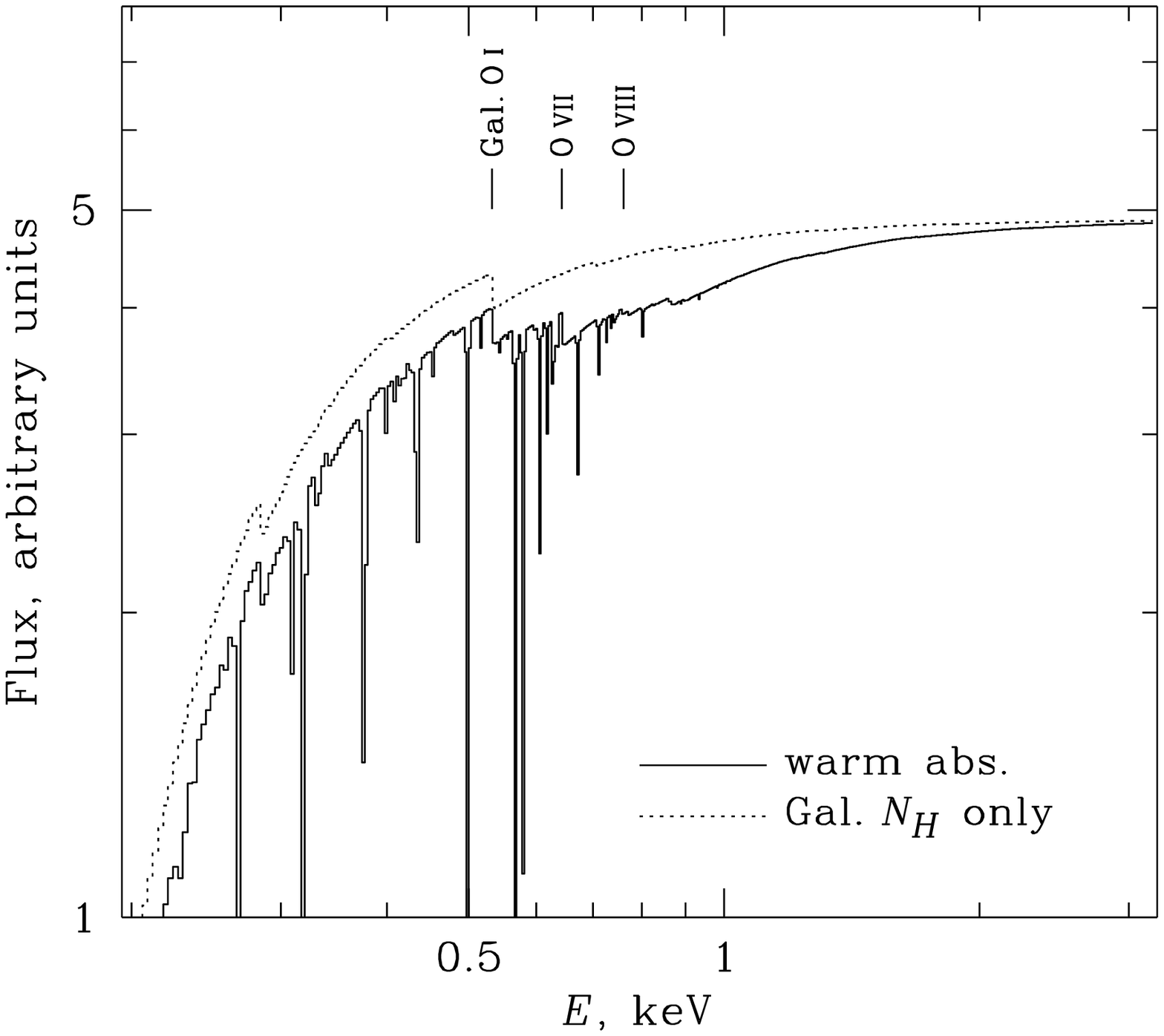}}

\rput[tl]{0}(9.5,20.7){\epsfxsize=8.5cm
\epsffile{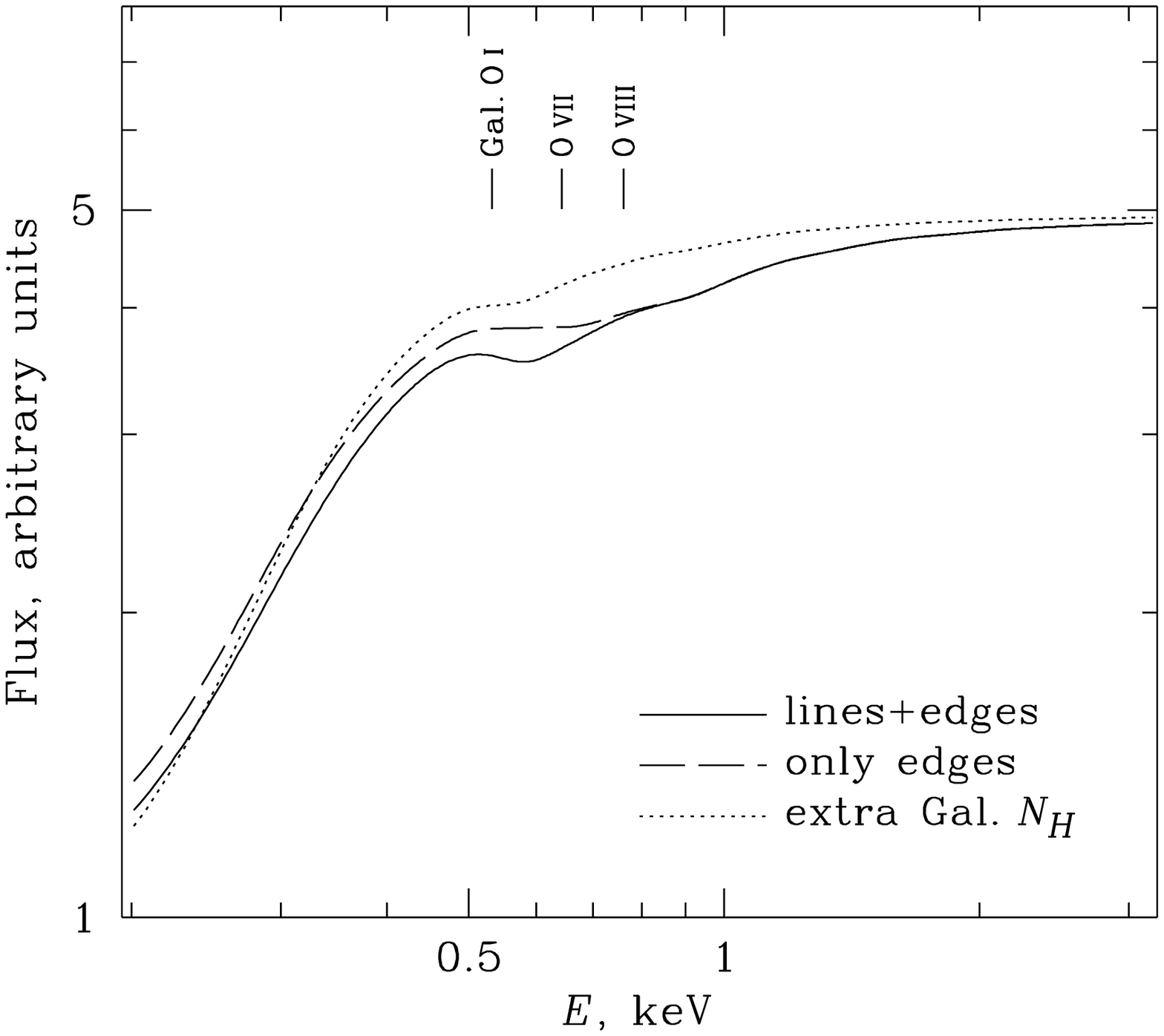}}

\rput[bl]{0}( 2.1,18.9){\large\it a}
\rput[bl]{0}(11.2,18.9){\large\it b}

\rput[tl]{0}(0,12.6){
\begin{minipage}{18cm}
\small\parindent=3.5mm
{\sc Fig.}~3.---({\em a}) Partially ionized absorber with $N_H=2\times
10^{21}$\cmsq\ and metallicity of 0.3 solar. Ion abundances are calculated
assuming $T=3\times 10^5$~K. For illustration, the absorber is placed at
$z=0.15$ although it will be spread over a redshift range $0.1-0.2$. The
absorption is applied to a flat spectrum with Galactic $N_H=2\times
10^{20}$\cmsq\ (dotted line). The spectral resolution is 3~eV. Energies of
$K\alpha$ edges of neutral Galactic oxygen and redshifted O{\small VII} and
O{\small VIII} are marked. ({\em b}) The same absorbed spectrum with a
resolution of 100~eV (solid line). Dashed line shows the absorption due to
edges only.  Dotted line now shows a spectrum without warm absorption but
with increased Galactic $N_H$ ($2.5\times 10^{20}$\cmsq) to match the flux
at $E\approx 0.3$ keV.
\end{minipage}
}
\endpspicture
\end{figure*}
%%%%%%%%%%%%%%%%%%%%%%%%%%%%%%%%%%%%%%%%%%%%%%%%%%%%%%%%%%%%%%%%%%%

\section{DISCUSSION}

\subsection{Absorption lines vs.\ absorption edges}
\label{sec:edges}

Earlier work addressing the practical possibility of detecting the IGM has
concentrated on resonant absorption lines in the spectra of distant quasars.
In the random direction in the sky, the column density of the absorbing
material is expected to be low and absorbers spread over a large redshift
interval. For such random searches, it is indeed optimal to look for easily
identifiable spectral features such as lines.  The lines are expected to
have very low equivalent width, $\Delta E \sim 0.1$ eV, and their detection
requires high spectral resolution instruments and large-area telescopes such
as the future calorimeter onboard {\em Constellation-X}%
\footnote{http://constellation.gsfc.nasa.gov}%
, or impractically long observations with grating spectrometers onboard the
forthcoming \chandra%
\footnote{http://asc.harvard.edu}
and \xmm%
\footnote{http://heasarc.gsfc.nasa.gov/docs/xmm}
observatories. 

It is often overlooked, however, that equivalent width of the
photoionization edges for oxygen ions is comparable to or greater than the
width of the resonant lines. Detection of an edge does not require high
spectral resolution. For example, Figure 3{\em b} shows the absorbed
spectrum from Fig.\ 3{\em a} smoothed with a 100 eV resolution (FWHM) that
can be achieved with a CCD. To emulate a realistic situation when the exact
Galactic column density is unknown and is fitted as a free parameter, the
figure also shows a model spectrum without the IGM absorption but with an
increased Galactic $N_H$ to reproduce the flux at low energies (dotted
line). Also shown by dashed line is an IGM-absorbed spectrum in which line
absorption is not included (only edges are included). It is apparent that
with such energy resolution, edge absorption in the interval $E=0.6-1$ keV
(observer frame) is a dominant feature. This absorption is ``warm'' and
cannot be mimicked by increasing the assumed neutral Galactic column
density.  Note, however, that because weak, broad edges are not so easily
identifiable as lines, one can hope to find them only if the redshift of the
absorber is known a priori, as in the Aquarius filament. In this filament,
the width of the redshift interval containing the absorber, $0.1\lax z\lax
0.2$, corresponds to an energy interval smaller than the spectral resolution
of the CCD detector and is therefore unimportant.

\subsection{Feasibility with forthcoming observatories}

The spectrum in Fig.\ 3 assumes an optimistic IGM column density expected
toward the most distant clusters in the Aquarius filament. Such absorption
can be detected already with the CCD instruments ACIS-S onboard \chandra\
and EPIC onboard \xmm, scheduled for launch in the near future. These
detectors will have the required sensitivity at the energies below $\sim0.4$
keV to straddle the expected absorption feature. Because this measurement
does not require high spectral resolution of grating spectrometers that have
limited efficiency, it can take advantage of the full effective area of the
telescopes, making the required exposures practical. Detailed simulations
show that for each of the several X-ray clusters on the far end of the
filament, a $3\sigma$ or better detection of the IGM absorption with the
above parameters can be obtained in a $\sim 6\times 10^4$~s observation with
\xmm\ or in a 2--3 times longer but still practical exposure with \chandra.
If the IGM is clumpy so that oxygen is in the lower ionization states that
have higher absorption depth, its detection would require shorter exposures.
Of course, if the unknown IGM density or its metallicity turn out to be much
lower than the above upper-limit estimates, longer exposures will be
required. For a differential test described in \S2, a significant number of
clusters--members of the filament needs to be studied.  As Fig.\ 2 shows,
there is a number of appropriate objects, and also an area with more
potential candidates with unknown redshifts. Thus the test proposed above
appears to be feasible with \chandra\ and \xmm, if the density and
metallicity of low-redshift IGM are close to the above optimistic
assumptions.

\subsection{Clusters as background candles}
\label{sec:clust}

Previous work emphasized quasars as background candles for the search of IGM
absorption. By studying distant, randomly selected quasars with future
large-area telescopes, one may eventually be able to determine the average
properties of IGM. At present, since IGM has not even been detected yet, one
can improve the chances of finding its traces by using galaxy clusters as
background sources. Clusters form at the intersection of matter filaments
(e.g., Colberg et al.\ 1997) which greatly increases the probability of a
favorable line of sight through a dense region of the universe. Because
clusters are extended, any absorption detected in their spectrum would have
to arise in the truly diffuse medium as opposed to a possible intervening
gas-rich galaxy or a Ly$\alpha$ cloud in front of a quasar. Although
clusters may exhibit intrinsic absorption in their central cooling flow
regions (e.g., Allen \& Fabian 1994), these regions can easily be masked out
from the spectral analysis with an imaging instrument. However, because the
angular extent of clusters precludes the use of grating spectrometers such
as those onboard \chandra\ and \xmm, they would require a calorimeter, such
as the future {\em Constellation X} or smaller-scale missions, to detect
weak absorption lines arising in the IGM with lower column densities than
that expected in the unique Aquarius filament.

\acknowledgements

The author is grateful to C. Otani for a conversation in 1994 in which the
idea of the proposed measurement has originated, and to M. Elvis, T.
Aldcroft, W. Forman, A. Vikhlinin, L. David and the referee for many useful
comments.  This work was supported by NASA contract NAS8-39073.


\begin{references}

\reference{} Abell, G. O. 1961, AJ, 66, 607

\reference{} Abell, G. O., Corwin, H. G., \& Olowin, R. P. 1989, ApJS, 70, 1

\reference{} Aldcroft, T., Elvis, M., McDowell, J., \& Fiore, F. 1994, ApJ,
437, 584 

\reference{} Allen, S. W., \& Fabian, A. C. 1994, MNRAS, 269, 409

\reference{} Burles, S., Tytler, D. 1996, ApJ, 460, 584

\reference{} Cen, R., \& Ostriker, J. P. 1999a, ApJ, 514, 1

\reference{} Cen, R., \& Ostriker, J. P. 1999b, preprint astro-ph/9903207
%(ApJ Letters in press)

\reference{} Ciardullo, R., Ford, H., \& Harms, R. 1985, ApJ, 293, 69

\reference{} Chen, L.-W., Fabian, A. C., \& Gendreau, K. C. 1997, MNRAS,
285, 449

\reference{} Colberg, J. M., White, S. D.M., Jenkins, A., \& Pearce, F. R.
1997, preprint astro-ph/9711040

\reference{} Colberg, J. M., et al.\ 1998, preprint astro-ph/9808257

%\reference{} David, L., Forman, W., \& Jones, C. 1991, ApJ, 380, 39

\reference{} Davidsen, A. F., Kriss, G. A., \& Wei, Zh. 1996, Nature, 380, 47

\reference{} De Young, D. S., 1978, ApJ, 223, 47

\reference{} Edge, A., \& Stewart, G. 1991, MNRAS, 252, 428

\reference{} Fang, T., \& Canizares, C. R. 1997, BAAS 191, 112206 

\reference{} Fukazawa, Y., Makishima, K., Tamura, T., Ezawa, H., Xu, 
H., Ikebe, Y., Kikuchi, K. \& Ohashi, T. 1998, PASJ, 50, 187 

\reference{} Fukugita, M., Hogan, C. J., \& Peebles, P. J. E. 1998, ApJ,
503, 518

\reference{} Giallongo, E., Fontana, A., \& Madau, P. 1997, MNRAS, 289, 629

\reference{} Gunn, J. E., \& Peterson, B. E. 1965, ApJ, 142, 1633

\reference{} Hellsten, U., Gnedin, N. Y., \& Miralda-Escud\'e, J. 1998, ApJ,
509, 56

\reference{} Kallman, T. R., \& Krolik, J. H. 1997, NASA/GSFC preprint (XSTA

\reference{} Miralda-Escud\'e, J., Cen, R., Ostriker, J. P., \& Rauch, M.
1996, ApJ, 471, 582 

\reference{} Mushotzky, R. F., \& Loewenstein, M. 1997, ApJ, 481, L63

%\reference{} Norman, C., \& Silk, J. 1979, ApJ, 233, L1

\reference{} Perna, R., \& Loeb, A. 1998, ApJ, 503, L135 

\reference{} Persic, M., \& Salucci, P. 1992, MNRAS, 258, 14p

\reference{} Rauch, M., et al.\ 1997, ApJ, 489, 7

\reference{} Renzini, A. 1999, astro-ph/9902361

\reference{} Sarazin, C. L. 1988, X-ray Emission from Clusters of Galaxies

\reference{} Shapiro, P. R., \& Bahcall, J. N. 1980, ApJ, 241, 1

\reference{} Steigman, G. 1996, preprint astro-ph/9610113

\reference{} Verner, D. A., Ferland, G. J., Korista, K. T. \& Yakovlev, D.
G. 1996a, ApJ, 465, 487

\reference{} Verner, D. A., Verner, E. M., \& Ferland, G. J. 1996b, Atomic
Data Nucl.\ Data Tables, in press

\reference{} Walker, T. P., Steigman, G., Kang, H.-S., Schramm, D. M.,
Olive, K. A. 1991, ApJ, 376, 51 

\end{references}
\end{document}